\documentclass[aps,pre,reprint,floatfix,superscriptaddress]{revtex4-2}

\usepackage{amsmath}
\usepackage{amssymb}
\usepackage{bm}
\usepackage{graphicx,subfigure}
\usepackage{epsf,epsfig}
\usepackage{hyperref}
\usepackage{xcolor}

\newcommand{\AI}[1]{\textcolor{black}{#1}}

\begin{document}

\title{The nonlinear motion of cells subject to external forces}

\author{Aondoyima Ioratim-Uba}
\affiliation{School of Mathematics, University of Bristol - Bristol BS8 1UG, UK}
\author{Aurore Loisy}
\affiliation{School of Mathematics, University of Bristol - Bristol BS8 1UG, UK}
\author{Silke Henkes}
\affiliation{School of Mathematics, University of Bristol - Bristol BS8 1UG, UK}
\affiliation{Lorentz Institute for Theoretical Physics, Leiden University - Leiden 2333 CA, The Netherlands}
\author{Tanniemola B. Liverpool}
\affiliation{School of Mathematics, University of Bristol - Bristol BS8 1UG, UK}

\date{\today}

\begin{abstract}
To develop a minimal model for a cell moving in a crowded environment such as in tissue, we investigate the response of a liquid drop of active matter moving on a flat rigid substrate to forces applied at its boundaries. We consider two different self-propulsion mechanisms, active stresses and treadmilling polymerisation, and we investigate how the active drop motion is altered by these surface forces. We find a highly non-linear response to forces that we characterise using drop velocity, drop shape, and the traction between the drop and the substrate. Each self-propulsion mechanism gives rise to two main modes of motion: a long thin drop with zero traction in the bulk, mostly occurring under strong stretching forces, and a parabolic drop with finite traction in the bulk, mostly occurring under strong squeezing forces. In each case there is a sharp transition between parabolic, and long thin drops as a function of the applied forces and indications of drop break-up where large forces stretch the drop.
\end{abstract}

\maketitle

\section{Introduction}

Cells are highly adaptable and move themselves around in a variety of different conditions and environments \cite{Mol_bio_of_the_cell, paluch2016focal,lammermann2009mechanical}. This is essential for biological functions such as wound repair \cite{chiou2018why}, organ development \cite{lecuit2007cell}, and in pathological processes such as cancer metastasis \cite{chaffer2011perspective}. Understanding individual cell motility and how it affects collective cell migration is key to understanding these processes. In particular, experiments show that cell-cell tugging plays an important role in collective migration \cite{trepat2009physical, jain2020role, tambe2011collective}, and that the force distribution within tissues may tell us something about pathological behaviour \cite{ng2014mapping,leal2017size}.

Cell motility is powered by the cytoskeleton, a dynamic network of interlinking protein filaments inside the cell \cite{mitchison1996actin, fletcher2010cell, juelicher2007active}. These filaments can collectively form anisotropic liquid crystalline (LC) phases~\cite{Marchetti2013}. There are a number of mechanisms by which cell motility occurs. 
 
The most studied is cell crawling \cite{barnhart2011adhesion, verkhovsky1999self, yam2007actin, keren2008mechanism, anon2012cell, wolgemuth2011redundant}, which combines the treadmilling (polymerisation/depolymerisation) of cytoskeletal actin filaments with strong adhesion to the substrate. In cells the likely source of this type of motion is actin polymerisation combined with acto-myosin contractility. Myosin II molecular motors cause actin filaments to slide relative to each other \cite{murrell2015forcing, pandya2017actomyosin} and generate an active stress that can be contractile (positive) or extensile (negative). A large class of hydrodynamic active liquid models of these proceses have been built, using active gel theory~\cite{callan2008viscous, recho2013asymmetry,blanch2013spontaneous, putelat2018mechanical, recho2019force}, and also more detailed computational active nematic LC models~\cite{tjhung2015minimal}. Such models can be augmented by including reaction-diffusion chemical feedback~\cite{camley2017crawling, doubrovinski2011cell}, and by adding confinement \cite{lavi2020motility, hawkins2009pushing}. It has been shown that even in the absence of actin treadmilling, spontaneous motion is still possible in LC active matter systems~\cite{tjhung2012spontaneous, khoromskaia2015motility, giomi2014spontaneous, hawkins2011spontaneous,sanchez2012spontaneous}.

\begin{figure*}[t]
	\centering
	\includegraphics[width=1.0\textwidth]{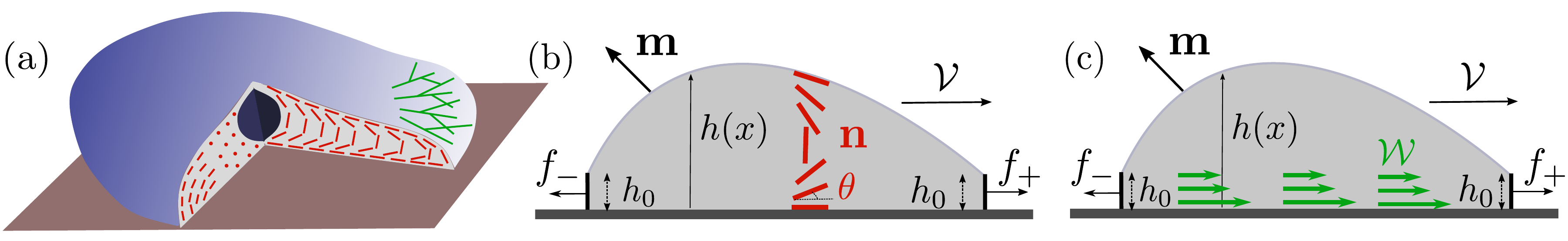}
	\caption{(a) Cartoon of a crawling cell with slice removed. The configuration of the actin filaments is shown in red, and polymerising filaments are shown in green. (b) Schematic showing a contractile/extensile drop on a flat surface being pushed/pulled at its boundaries by external forces $f_{\pm}$ (which can also stretch or squeeze the drop), positive $f_{+}$ or $f_{-}$ means that the force is directed to the right. The nematic director is denoted by $\mathbf{n}$. (c) Schematic showing a polymerising drop with external forces $f_{\pm}$. In both cases, the boundaries are set to be at a height $h_{0}$ above the solid substrate ($z=0$).}
	\label{fig:CellDiagram}
\end{figure*}

A minimal description of cell motility is thus provided by the motion of a drop of anisotropic active LC matter. 
The possible modes of a such a drop freely moving on a flat rigid substrate have recently been classified by some of us in \cite{loisy2020modes}. We identified three modes: motion due to active stresses, motion due to self-advection of active units along their direction of orientation, and motion due to contact angle mismatch. All modes are in principle present but one can consider regimes when one mode dominates. We also showed that a drop moving purely due to active stresses can do so without exerting traction on the substrate \cite{loisy2019tractionless}. This type of motion is relevant to fast migration in crowded environments \cite{paluch2016focal,liu2015confinement}, where cellular adhesions are unstable at high strain rates \cite{schwarz2013adherentCells}. 

Crowded environments also lead to significant forces on cells.
However, many of the effects of external forces on cell-motility and migration remain a mystery. These forces, either coming from cell-cell tugging or from outside the cell, are important for the functionality of cells and tissues. Furthermore, experiments show that external forces can alter cell stiffness, induce migration, alter cell shape, induce remodelling, and alter cell phenotype~\cite{li2019application, saez2007rigidity, ricca2018transient}. Hence a better handle on them promises to have a significant impact on our understanding of multicellular systems.

In this paper we study the dynamics of an active LC drop on a flat surface under external forces applied at its two ends. We model the anisotropy of the cytoskeleton in two ways, corresponding to \AI{dynamics dominated by} the first two modes we identified in \cite{loisy2020modes}. First, we probe an imposed LC director field, and the active stress that this generates (active contractile/extensile drop). The behaviour is controlled by the ratio $\mathcal{A}$ of the activity to the splay-bend winding number of the LC director, as measured from the surface to the top of the drop. Second, we study self-advection of LC units along their direction of orientation (active polymerising drop), whose behaviour is controlled by the ratio $\mathcal{W}$ of the self-advection speed to the surface tension. We classify the motion of the drop according to (1) the difference between the forces on each end, i.e. whether it is being squeezed or stretched and (2) the sum of the forces applied to its two ends, i.e. if it is being pushed to the right (R) or to the left (L). For a passive drop ($\mathcal{A}=0$, $\mathcal{W} = 0$), we find parabolic shapes with simple symmetric behaviour: it moves to the right (or left) if it is pushed in the right (or left) direction unless it is stretched above a critical stretching force where it tends to break up into smaller droplets. For both active drops, we find a much richer response to external forces. The active contractile/extensile drop ($\mathcal{A}>0$, $\mathcal{W}=0$) drop moves to the R almost all the time except when the sum of forces is large in the L direction and it is being squeezed.  We also find a wide variety of shapes. Parabolic shapes are observed only when the the drop is being squeezed. When the drop is being stretched, we find (i) a double humped shape with a large hump at the front when the sum of forces is strongly to the R, (ii) a flat pancake shape which exerts almost no traction on the surface  when the sum of forces is small and (iii) droplet break up when the sum of forces is strongly in the  L direction. Our results hold for both contractile and extensile drops as our equations of motion are invariant: changing the sign of  $\mathcal{A}$, which flips the directions L $\leftrightarrow$ R, is equivalent to either moving from contractile to extensile stresses or to flipping the splay-bend winding number. The active polymerising drop ($\mathcal{A}=0$, $\mathcal{W}>0$) also moves to the R almost all the time, except when the sum of forces is large in the L direction and it is being squeezed. Again, parabolic drops are observed only when the drop is being squeezed. When the drop is being stretched, we find (i) a double humped shape with a large hump at the rear when the sum of forces is strongly to the L, (ii) a flat travelator shape which exerts almost no traction on the surface and (iii) droplet break up when the sum of forces is strongly in the R direction. Changing the sign of  $\mathcal{W}$ flips the directions L $\leftrightarrow$ R. 

\begin{figure*}[t]
	\centering
	\includegraphics[width=1.0\textwidth]{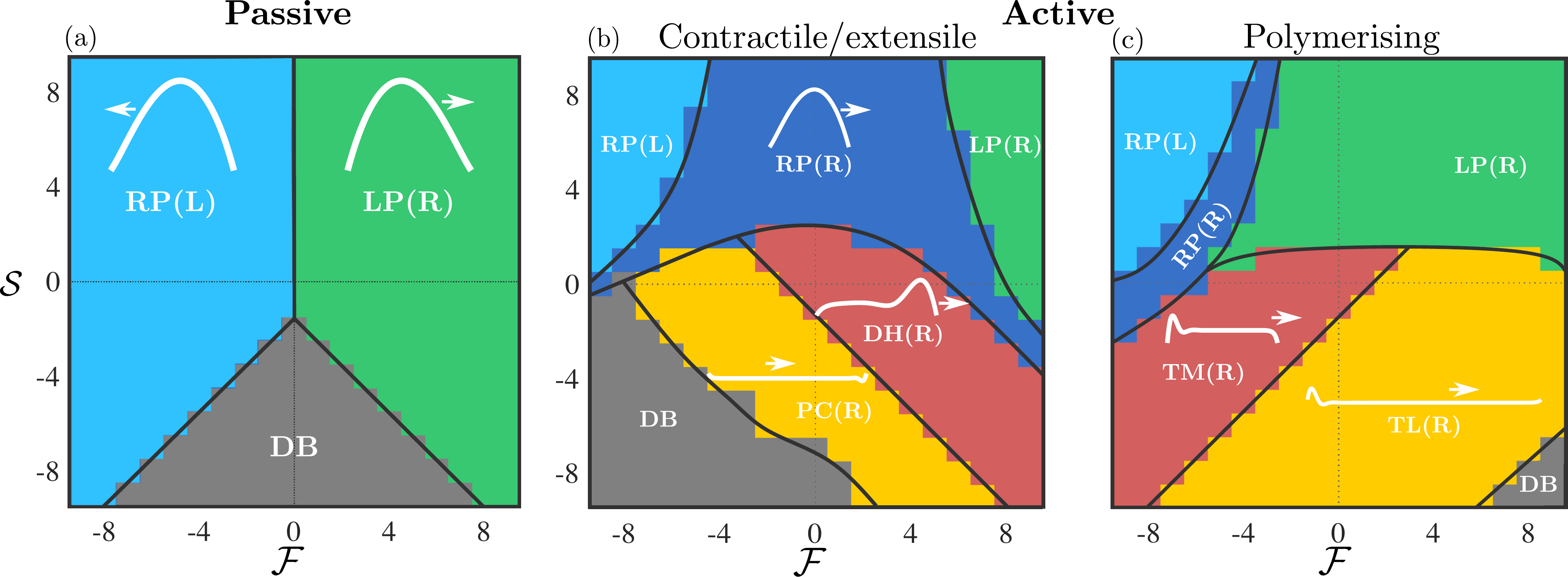}
	\caption{(a) Schematic phase diagram for a passive drop. The light blue region corresponds to RP(L) right-parabola left-moving drops, green corresponds to LP(R) left-parabola right-moving drops, while droplet breakup DB is grey. (b) Schematic phase diagram for a contractile/extensile drop. New phases include the RP(R) right-parabola right-moving state (dark blue), the DH(R) double-hump right-moving (coral) and the PC(R) pancake right-moving (yellow) states. (c) Schematic phase diagram for a polymerising drop. The RP(L), LP(R), and RP(R) phases are present, while the DH(R) double-hump, right-moving is replaced by the TM(R) tread-milling drop right-moving (coral), and the PC(R) pancake right-moving is replaced by the TL(R) travelator right-moving (yellow). Characteristic drop shapes are superimposed in white.}
	\label{fig:schematic}
\end{figure*}

\section{Model}
We model a single cell as a two dimensional incompressible active nematic drop moving on a flat rigid substrate subject to external forces at its boundaries (Fig. \ref{fig:CellDiagram} (a)). This can be thought of as a projection of the full 3d system to its average in 2d. \AI{We expect this minimal model to capture a large part of the phenomenology in 3d but not the full behaviour such as splay in the lamellipodial protrusion \cite{tjhung2015minimal}.} We work in the $(\tilde{x},\tilde{z})$ plane, where the drop is characterised by the height $\tilde{h}(\tilde{x},\tilde{t})$ of its free surface above the substrate, and moves with velocity $\tilde{V}$. \AI{Note that variables with tildes are dimension-full, while the un-tilded variables, which we introduce for the lubrication approximation \cite{oron1997long}, are dimension-less.} We use the well established equations of active liquid crystal hydrodynamics \cite{marchetti2013hydrodynamics, julicher2018hydrodynamic} where the motion of the coarse grained orientation of elongated units $\mathbf{n} = (\cos\theta, \sin\theta)$, known as the director, is coupled to the fluid velocity $\tilde{\mathbf{u}}$ inside the drop. The velocity satisfies force balance equations at vanishing Reynolds number:
\begin{subequations}
\begin{equation} \label{eq:forceBalance}
\partial_{j}\tilde{\sigma}_{ij} + \tilde{f}_{i} = 0,
\end{equation}
\begin{equation}
\tilde{\sigma}_{ij} = -\tilde{p}\delta_{ij} + \eta(\partial_{i}\tilde{u}_{j} + \partial_{j}\tilde{u}_{i}) - \tilde{\alpha} n_{i}n_{j},
\end{equation}
\end{subequations}
where $\tilde{\mathbf{f}} = \tilde{\mathbf{f}}(\tilde{x})$ is the external force per unit height, $\tilde{p}$ is the pressure, $\tilde{\mathbf{u}}$ is the fluid velocity inside the drop, and $\tilde{\alpha}\AI{n_in_j}$ is the active stress \AI{to leading order in a gradient expansion}, which represents the coarse-grained stresses generated when cytoskeletal actin filaments slide relative to each other \cite{hatwalne2004rheology,liverpool2006rheology}. \AI{Following \cite{loisy2019tractionless,loisy2020modes}, we perform calculations in the lubrication approximation \cite{oron1997long} for which changes in the height are much smaller than the width. In this approximation, higher order gradients in the stress tensor can be neglected because they scale with the ratio of the characteristic length to the characteristic height of the drop, which is small. We also work in the strong elastic limit, where the director relaxes instantaneously to follow changes in the drop height $\tilde{h}(\tilde{x},\tilde{t})$. In this case, the dynamic equation for the director reduces to $\mathbf{\nabla}^2\theta = 0$.} 

For the active contractile/extensile drop, we anchor the director parallel to the substrate, i.e. $\theta(\tilde{z} = 0) = 0$, and parallel to the free surface with an imposed winding, i.e. $\theta(\tilde{z} = \tilde{h}) = \omega\pi + \arctan(h^{\prime})$. The winding number $\omega \in \mathbb{Z}^{+}$ counts the number of half turns of the director across the drop height (Fig. \ref{fig:schematic}). At fixed activity, the winding of the director, through its chirality, breaks the left-right symmetry and sets the preferred direction of motion. With no external forces, at positive activity (extensile drop), the drop moves right for positive winding and left for negative winding. There is no active contribution to the drop velocity when $\omega = 0$. For the active polymerising drop, we anchor the director parallel to the substrate i.e. $\theta(\tilde{z} = 0) = 0$, and parallel to the free surface with no imposed winding, i.e. $\theta(\tilde{z} = \tilde{h}) =\arctan(h^{\prime})$.

The external forces $\tilde{f}_{\tilde{x}} = \tilde{f}_{+}\delta(\tilde{x}-\tfrac{\tilde{L}}{2}) + \tilde{f}_{-}\delta(\tilde{x}+\tfrac{\tilde{L}}{2})$, $\tilde{f}_{\tilde{z}} = 0$ are localised at the left and right drop borders, which leads to boundary conditions on the pressure $\tilde{p}(\pm \tilde{L}/2) = \tilde{\pi}_{0} \mp \tilde{f}_{\pm}$, and $\tilde{\pi}_{0} = 2\gamma\tilde{\phi}/\tilde{L}_{0}$ is the Laplace pressure generated by the surface tension $\gamma$ of an \AI{un-forced} symmetric passive drop of length $\tilde{L}_{0}$ and contact angle $\tilde{\phi}$. For an active drop with no external forces, $\tilde{\pi}_{0}$ is a constant shift in the pressure. These boundary conditions can be derived by integrating the $\tilde{x}$ component of \eqref{eq:forceBalance} with respect to $\tilde{x}$ from $\tilde{L}/2 - \Delta$ to $\tilde{L}/2 + \Delta$ and from $-\tilde{L}/2 - \Delta$ to $-\tilde{L}/2 + \Delta$, \AI{where $\tilde{L}$ is the length of an arbitrary drop}, with $\Delta > 0$, and taking the limit $\Delta \to 0^{+}$ (\AI{See section 1.2 of SI \cite{SI}}). In addition to these boundary conditions, we have $\tilde{\bm{\sigma}}\cdot\bm{m} = \gamma\kappa\bm{m}$ for the stress at the free surface, where $\gamma$ is the surface tension, $\bm{m}$ is the unit normal vector (see Fig 1(a)), and $\kappa = \bm{\nabla}\cdot\bm{m}$ is the curvature of the free surface. The interaction of the drop with the rigid substrate is modelled by a partial slip boundary condition: $\tilde{u}_{\tilde{x}} = \tilde{l}_{u}\tilde{\sigma}_{\tilde{x}\tilde{z}}/\eta$, where $\tilde{l}_{u}$ is a slip length. 

We seek travelling wave solutions $\tilde{h} = \tilde{h}(\tilde{x}-\tilde{V}t)$ \AI{and work in the reference frame of the drop. Mass conservation implies}
\begin{equation} \label{eq:mass_conservation}
\int_{0}^{\tilde{h}} (\tilde{u}_{\tilde{x}} + \tilde{w}n_{x} + \tilde{V}) \,d\tilde{z} = 0,
\end{equation}
where $\tilde{\bm{u}}$ is the fluid velocity inside the drop satisfying force balance and incompressibility, and $\tilde{w}\bm{n}$ describes the additional transport due to treadmilling self-advection at speed $\tilde{w}$ of active units whose orientations are characterised by the director $\bm{n}$ - the active units essentially propel themselves along their own tangent. Since treadmilling \AI{self-advection} only occurs near substrates, we choose the form  
$\tilde{w} = \tilde{w}_{0}\:e^{-\tilde{h}/\tilde{l}_{w}}$,
where $\tilde{w}_0$ is a characteristic self-advection speed and $\tilde{l}_{w}$ is the characteristic height over which the self-advection term decays in the direction normal to the substrate. This functional form was also used in previous work that simulates crawling cells \cite{tjhung2015minimal}. Then, combining mass conservation with $\tilde{p} = -\gamma\tilde{h}^{\prime\prime}$, which can be derived by taking the leading order terms of the normal component of the stress boundary condition at the drop free surface, we obtain a non-linear ODE for $\tilde{h}$. 
\AI{After non-dimensionalisation, we retain the salient dimensionless parameters activity $\mathcal{A}$, advection speed $\mathcal{W}$, and drop velocity $\mathcal{V}$,
\begin{equation} \label{eq:NonDimParameters}
\mathcal{A} = \frac{\tilde{\alpha} \tilde{L}_{x}}{4\pi\omega\gamma\epsilon^2},  \quad  \mathcal{W} = \frac{\eta \tilde{w}_{0}}{\gamma\epsilon^3}, \quad  \mathcal{V} = \frac{\eta \tilde{V}}{\gamma\epsilon^3}
\end{equation}
and also height $h = \tilde{h}/\epsilon \tilde{L}_{x}$, coordinates $x = \tilde{x}/\tilde{L}_{x}$, $z = \tilde{z}/\epsilon \tilde{L}_{x}$, slip length $l_{u} = \tilde{l}_{u}/\epsilon \tilde{L}_{x}$, polymerisation height $l_{w} = \tilde{l}_{w}/\epsilon \tilde{L}_{x}$, pressure $\pi_{0} = \tilde{\pi}_{0}\tilde{L}_{x}/\gamma\epsilon$, contact angle $\phi = \tilde{\phi}/\epsilon$, and forces $f_{\pm} = \tilde{f}_{\pm}\tilde{L}_{x}/\gamma\epsilon$, where $\tilde{L}_{x}$ is a characteristic length scale in the $x$ direction.} Then the drop shape satisfies the ODE
\begin{equation} \label{eq:ODE}
h^{\prime\prime\prime}\bigg(\frac{1}{3}h^3 + l_{u}h^2\bigg) + g(h) = \mathcal{V}h,
\end{equation}
where $g(h) = \mathcal{A}h^2$ for the contractile/extensile drop and $g(h) = \mathcal{W}l_w(1 -\exp(-h/l_w))$ for the polymerising drop. The drop velocity $\mathcal{V}$ is a functional of $h$:
\begin{equation} \label{eq:dropVelocity}
\mathcal{V} = \left[ \int_{\tfrac{-L}{2}}^{\tfrac{L}{2}} \frac{g(h) \: dx}{\tfrac{h^3}{3} + l_{u}h^2} + \mathcal{F} \right] \bigg/ \left[\int_{\tfrac{-L}{2}}^{\tfrac{L}{2}} \frac{dx}{\tfrac{h^2}{3} + l_{u}h} \right],
\end{equation}
where we have introduced push $\mathcal{F} = f_{+} + f_{-}$ (positive means push right and negative means push left). We also use the quantity $\mathcal{S} = -( f_{+} - f_{-})$, which we term squeeze, in the analysis of our results (positive means squeeze, and negative means stretch). We set the height of the drop at its boundary to a finite $h_0 = h(\pm L/2)$, which represents a finite contact area with neighbouring cells or with obstacles. The drop length $L$ is determined as part of the solution by requiring that the drop have constant volume $\Omega$: $\int_{-L/2}^{L/2} h dx = \Omega$. The tangential traction exerted by the drop on the substrate is $\sigma_{xz}|_{z = 0} = hh^{\prime\prime\prime}$. In terms of $h$ alone, using \eqref{eq:ODE}, the traction can be written $\sigma_{xz}|_{z = 0} = (\mathcal{V} - g(h)/h)/(\tfrac{1}{3}h + l_{u})$. The dimensionless parameters $\mathcal{A}$, $\mathcal{W}$, $\mathcal{V}$, $\mathcal{F}$, and $\mathcal{S}$ all represent a ratio of stress to surface tension, and as we will show below, the interesting active phenomenology appears when they are all $\mathcal{O}(1)$. We translate these results back to biological parameters at the end, allowing us to make predictions for cell speeds and stresses in addition to cell shapes.

\begin{figure}[t!]
	\centering
	\includegraphics[width=0.5\textwidth]{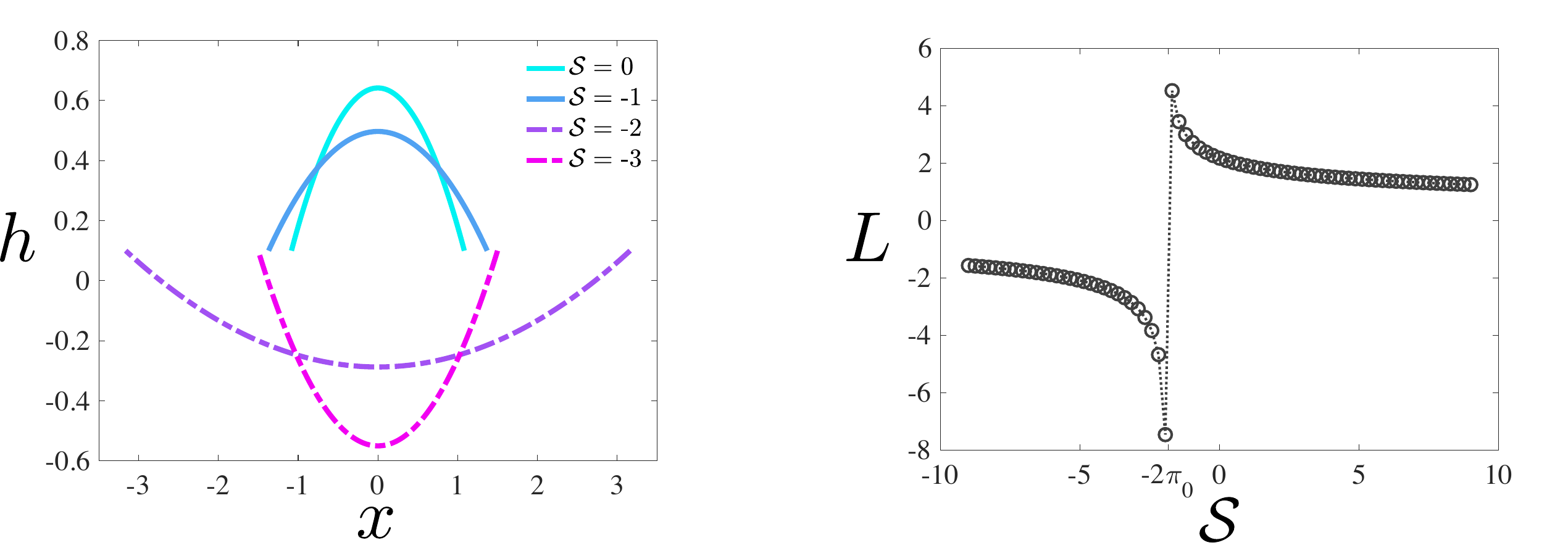}
	\caption{Left: droplet height profiles at $\mathcal{F} = 0$ and $\mathcal{A} = 0$. Here, $\pi_{0} = 0.9226$ and $\Omega = 1$. Right: drop length $L$ as a function of $\mathcal{S}$.}
	\label{fig:symmBreak}
\end{figure}

\subsection{Numerics}
We obtained solutions to \eqref{eq:ODE} by numerically solving the time dependent version of the force balance equation, $\partial_{t}h + \partial_{x}I = 0$, where $I = h^{\prime\prime\prime}(\tfrac{1}{3}h^3 + l_{u}h^2) + g(h) - \mathcal{V}h $, which represents overdamped dynamics of the height evolution. Solving the time dependent problem (starting from a known initial condition) allows us to iteratively evaluate the integrals in equation \eqref{eq:dropVelocity} and gives us the potential to investigate the temporal dynamics of the drop. The steady state of the time evolution, $\partial_{t}h = 0$, is equivalent to equation \eqref{eq:ODE}.

We used an implicit method, a variant of the Crank-Nicholson algorithm, using finite difference coefficients to approximate the derivatives. The algorithm starts with with an initial condition that has the correct values of $h$ at the boundaries and advances in time until steady state. The algorithm advances in time according to $h_i^{n+1}/\Delta t + I_i(h_1^{n+1}, \cdots, h_{N}^{n+1})/2 = h_i^{n}/\Delta t + I_i(h_1^{n}, \cdots, h_{N}^{n})/2$, where $h_{i}^{n}$ is the discretised drop height evaluated at the $i^{\text{th}}$ spatial point at the $n^{\text{th}}$ time step, $I_i(h_1^{n}, \cdots, h_{N}^{n})$ is the discretised vesrion of $I$ as defined above evaluated at the $i^{\text{th}}$ at the $n^{\text{th}}$ time step. The time stepping in the algorithm is adaptive, meaning that the time step is increased as steady state approaches \cite{loisy2020modes}. In the numerics, we apply both height and pressure boundary conditions (converted to boundary conditions on $h^{\prime\prime}$) because the time evolution is a fourth order PDE:
\begin{equation} \label{eq:BCs}
h(\pm L/2) = h_{0}, \quad h^{\prime\prime}(\pm L/2) = \pm f_{\pm} - \pi_{0}.
\end{equation}
The boundary conditions on $h^{\prime\prime}$ are required for consistency with force balance, however we are in principle free to change the boundary conditions on $h$. In all simulations $l_{u} = 0.05$, $h_{0} = 0.1$, $\Omega = 1$, $\pi_{0} = 0.9226$, which corresponds to a contact angle of $1$ for a free passive drop, $\omega = 1$, and the spatial step size is $0.002$. \AI{The value of the slip length was chosen to be consistent with \cite{loisy2020modes}. Changing the slip length has different effects on the contractile/extensile mode and polymerisation mode: increasing the slip parameter at a fixed polymerisation speed decreases drop speed because motility via polymerisation requires strong adhesion \cite{tjhung2015minimal,loisy2020modes}. The opposite is true for the contractile/extensile drops - increasing the slip length increases drop speed \cite{loisy2020modes}.}

\begin{figure*}[t]
	\centering
	\includegraphics[width=1.0\textwidth]{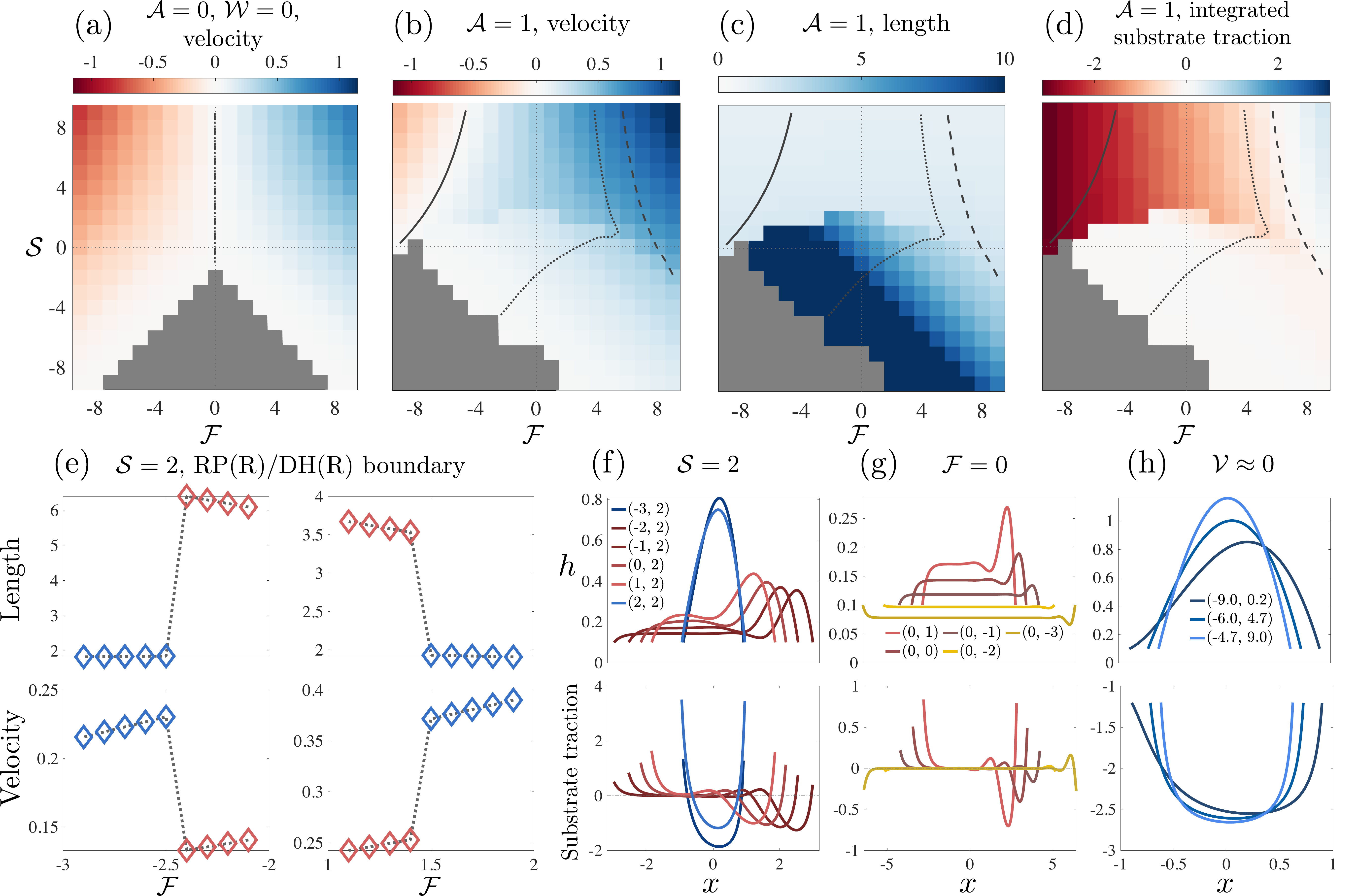}
	\caption{Contractile / extensile drops. (a) Velocity of a passive drop. (b) Velocity of a contractile/extensile drop. (c) Length of a contractile/extensile drop. (d) Traction between the drop and the substrate integrated over $70\%$ of the drop length: \AI{$\int \sigma_{xz}|_{z=0} \,dx$}. Each square represents a single simulation and its colour corresponds to the legend. The solid line in each plot is an isoline corresponding to zero velocity, the dashed line is an isoline where the first moment of $h$, $\int_{0}^{L} xh \,dx$, vanishes, and the dotted line is an isoline corresponding to the drop having equal contact angles. 
(e) 2\textsuperscript{nd} moment and velocity $\mathcal{V}$ across the RP(R)/RH(R) boundary, indicating a first order transition.
(f) Stable numerical solutions for $h$ (top) and the corresponding traction $\sigma_{xz}|_{z=0}$ (bottom) for $\mathcal{A} = 1$, $\mathcal{S} = 2$, (g) $\mathcal{A} = 1$, $\mathcal{F} = 0$, (h) $\mathcal{V} < 2 \times 10^{-4}$. The colour of each curve matches the colour of its phase in Fig. \ref{fig:schematic} (c). The legend labels are coordinates in the $(\mathcal{F},\mathcal{S})$ phase plane.}
	\label{fig:profiles}
\end{figure*}

\begin{figure*}[t]
	\centering
	\includegraphics[width=1.0\textwidth]{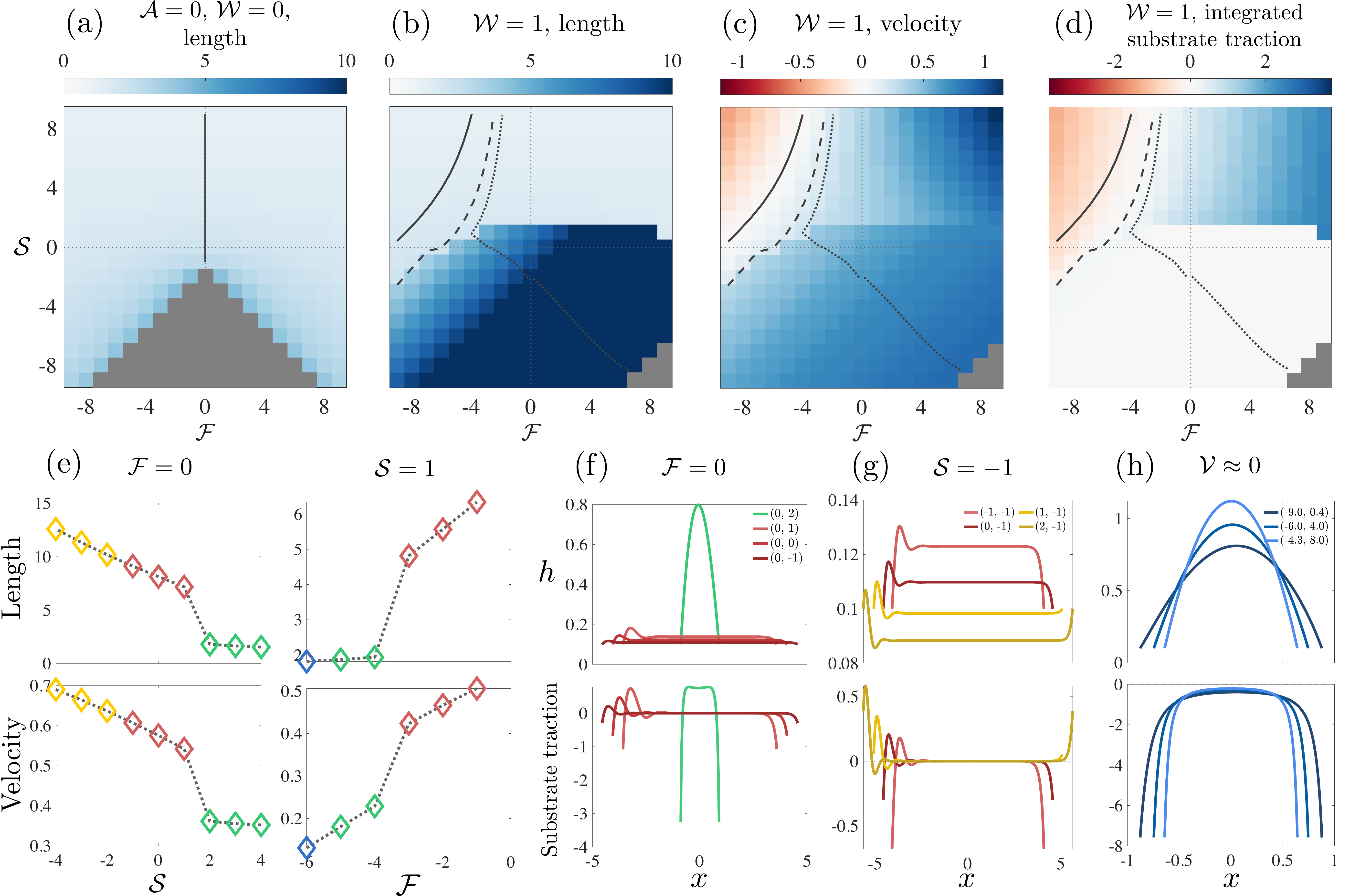}
	\caption{Polymerising drops. (a) Length of a passive drop. (b) Length of a polymerising drop. (c) Velocity of a polymerising drop. (d) Traction between the drop and the substrate integrated over $70\%$ of the drop length: \AI{$\int \sigma_{xz}|_{z=0} \,dx$}. Each square represents a single simulation and its colour corresponds to the legend. The solid line in each plot is an isoline corresponding to zero velocity, the dashed line is an isoline where the first moment of $h$, $\int_{0}^{L} xh \,dx$, vanishes, and the dotted line is an isoline corresponding to the drop having equal contact angles. (e) Jump in length and velocity going from parabolic to flat. (f) Stable numerical solutions for $h$ (top) and the corresponding traction $\sigma_{xz}|_{z=0}$ (bottom) for $\mathcal{W} = 1$, $\mathcal{F} = 0$, (g) Stable numerical solutions for $h$ (top) and the corresponding traction $\sigma_{xz}|_{z=0}$ (bottom) for $\mathcal{W} = 1$, $\mathcal{S} = -1$, (h) $\mathcal{V} < 2 \times 10^{-4}$. The colour of each curve matches the colour of its phase in Fig. \ref{fig:schematic} (c). The legend labels are coordinates in the $(\mathcal{F},\mathcal{S})$ phase plane.}
	\label{fig:profilesA}
\end{figure*}

\section{Results}

The phase diagrams in Fig. \ref{fig:schematic}a-c schematically show the regimes of drop shapes and motility that we find for the passive $\mathcal{A} = 0, \mathcal{W} = 0$, the contractile/extensile case $\mathcal{A} = 1, \mathcal{W} = 0$ case, and the polymerising case $\mathcal{A} = 0, \mathcal{W} = 1$. Figures \ref{fig:profiles} and \ref{fig:profilesA} provide full results for drop velocity, drop shapes and substrate traction. 

\subsection{Passive drop}
We begin with a passive drop $\mathcal{A} = 0, \mathcal{W}=0$ (Fig. \ref{fig:schematic} (a)), both to provide a reference, but also to illuminate the limit when surface tension dominates over activity. Then all dimensionless parameters are small and equation \eqref{eq:ODE} can be approximated by $h^{\prime\prime\prime}(x) = 0$, which is the equation for the passive drop with a parabolic solution. Moving to finite $\mathcal{F}$ and $\mathcal{S}$ now, we observe two phases of motion, both with finite traction in the bulk: a left-moving right-parabola RP(L) for $\mathcal{F} < 0$ and a right-moving left-parabola LP(R) for $\mathcal{F} > 0$. These passive drops are qualitatively close in shape to an upside down parabola but are asymmetric with a first moment of $h$ that is non-zero (characteristic shapes shown in Fig. \ref{fig:schematic} (b)). Squeezing the drop $\mathcal{S} > 0$ in general maintains the parabolic shape and increases the effective surface tension.

The passive solutions are antisymmetric along the push/pull $\mathcal{F}$ axis, and they move with opposite velocities that increase as the drop is squeezed (see Fig. \ref{fig:profiles}(a)).  All drops become longer and thinner as we move along the stretch/squeeze axis from squeeze to stretch, eventually reaching the region of drop breakup (DB), where the drop free surface reaches $h < 0$. For passive drops at $\mathcal{F}=0$, there are no steady solutions in the DB region that satisfy volume conservation. Consider a passive drop with $\mathcal{A} = 0$ and $\mathcal{F} = 0$. The equation for drop height becomes
\begin{equation}
h^{\prime\prime\prime}(\tfrac{1}{3}h^3 + l_{u}h^2) = 0
\end{equation}
and has the solution
\begin{equation}
h = -\frac{1}{2}\bigg(\pi_{0} + \frac{\mathcal{S}}{2}\bigg)\bigg(x^2 - \frac{L^2}{4}\bigg) + h_{0},
\end{equation}
where we have used $h(\pm L/2) = h_{0}$ and $h^{\prime\prime} = f_{+} - \pi_{0}$. The drop length $L$ is determined by the volume constraint $\int_{-L/2}^{L/2} h \,dx = \Omega > 0$. The drop length diverges at $\mathcal{S} = -2\pi_{0} \approx -1.99$ ($\pi_{0}$ is chosen so that the contact angle of a free passive drop is one), and goes negative for $\mathcal{S}  < -2\pi_{0}$. This means that volume conservation is not truly satisfied even though the condition $\int_{-L/2}^{L/2} h \,dx = \Omega$ is satisfied mathematically by a negative $L$. This is clear in the left plot in figure \ref{fig:symmBreak}, where $h < 0$ for $\mathcal{S}  < -2\pi_{0}$ and thus $\int_{-\mathopen|L\mathclose|/2}^{\mathopen|L\mathclose|/2} h \,dx < 0$, with $\mathopen|L\mathclose|$ being the actual length of the drop. 

\subsection{Active contractile/extensile drop}
With contractility, as shown in shown in Fig. \ref{fig:schematic} (b), we find, in addition to RP(L), LP(R) and DB, a right-moving right-parabola RP(R) state with finite traction, and then two long and thin states (see length in Figs. \ref{fig:profiles} (c)), the right-moving double hump DH(R) state, which has a small dip followed by a prominent frontal hump, and the right-moving pancake PC(R) drop, which is flat almost everywhere with the average height $\bar{h} < h_{0}$. 

Both DH(R) and PC(R) drops have zero traction in the bulk (Fig. \ref{fig:profiles} (d)). In \cite{loisy2019tractionless}, equal contact angles were imposed at the boundaries for all drops (rather than variable external force), and it was found that the tractionless DH(R) shapes emerged when the dimensionless active contractile stress  $\mathcal{A} \gtrsim 0.8$. Here, these tractionless drops can be found in the DH(R) region of the phase plane along the line of equal contact angle (dotted line, Figs. \ref{fig:profiles} (b) - (d)). The drop partitions itself into a small region with large free surface curvature, and a larger region where the free surface is flat. There is large curvature at these humps, and the remaining $50\% - 90\%$ of the drop is flat. At lower activity $\mathcal{A} = 0.3$, $\mathcal{A} = 0.1$ (see SI \cite{SI} Fig. S2), we find DH(R) and PC(R) shapes at large enough stretch ($\mathcal{S} < 0$), which shows that stretching can work together with activity to deform the drop. Therefore, in the DH(R) and PC(R) drops, we see the competition between active contractile/extensile stress plus stretch, and surface tension: activity plus stretch wants to deform the drop, and surface tension wants the free surface to be flat.

The symmetry about the push/pull axis is broken because the drop is motile at zero force due to the imposed winding $\omega = 1$ (Figs. \ref{fig:profiles} (b)). Both the RP(L) and LP(R) regions become smaller with activity and are pushed towards the top-left (push-left/squeeze) and top-right (push-right/squeeze) of the phase plane respectively by the emergence of the RP(R) phase, which largely occupies the middle squeeze region that lies between RP(L) and LP(R). The stretch region of the phase plane is populated by DH(R) and PC(R), as well as DB. The region containing DH(R) and PC(R) shares a phase boundary with RP(R) determined by the drop length, and the distribution of substrate traction. The second moment of $h$ (with drop length scaled to unity), which is a measure of the spread of the drop also determines the same phase boundary - see SI. The phase boundary is indicated in the schematic Fig. \ref{fig:schematic}, as determined from the data in Figs. \ref{fig:profiles} (c) and (d). Strikingly, the transition between RP(R) and DH(R)/PC(R) is sharp suggesting a first order transition. 

The behaviour of the second moment and the drop velocity across the phase boundary is shown in Fig. \ref{fig:profiles} (e). There is a jump in both quantities going from RP(R) to DH(R) and also from DH(R) to RP(R). The change in drop shape induced by moving from RP(R) to DH(R) and back to RP(R) at constant squeeze $\mathcal{S} = 2$ is shown if Fig. \ref{fig:profiles} (f). The drop shape changes continuously within the DH(R) region but changes sharply at the phase boundary. There are also indications of bi-stability in the region: we have obtained two stable solutions for two different initial conditions at the same point in the phase plane (see SI at \cite{SI} Fig.3). In contrast, the transition between DH(R) and PC(R) is smooth and happens under increased stretching for constant push (Fig. \ref{fig:profiles} (g)). 

\subsection{Active polymerising drop}
The polymerising drop, in addition to RP(L), LP(R), RP(R), and DB has two long and thin states (see length in Fig, \ref{fig:profilesA} (b)), the right-moving treadmilling state TM(R), which has a prominent hump at the rear followed by a small dip and a protrusion at the front, and the right-moving travelator, TL(R) which is flat almost everywhere, with the average height $\bar{h} < h_{0}$. The TM(R) shape is a solution to the 2D equation for a free drop driven by actin polymerisation for large enough self-advection velocity \cite{loisy2020modes}. Similar shapes with flat frontal protrusions have been observed experimentally for motile keratocytes \cite{keren2008mechanism, barnhart2011adhesion}. Both TM(R) and TL(R) drops have zero traction in the bulk \ref{fig:profilesA} (d). In \cite{loisy2020modes}, where equal contact angles (rather than variable force) were imposed, the TM(R) shapes emerged when the dimensionless polymerisation speed $\mathcal{W} \gtrsim 0.2$, showing that relatively small polymerisation speeds can generate stresses large enough to deform the drop. Again, there is large curvature at the humps, which take up $10\% - 50\%$ of the drop length, and the remaining drop is flat. At much lower polymerisation speed, $\mathcal{W} = 0.01$, (see SI \cite{SI}), we find TM(R) and TL(R) shapes at large enough stretch ($\mathcal{S} < 0$), which again shows that that stretching can amplify the effect of activity to deform the drop. Therefore, in the TM(R) and TL(R) drops, we see the competition between stress generated by polymerisation plus stretch, and surface tension: activity plus stretch wants to deform the drop, and surface tension wants the free surface to be flat.

Again, the symmetry about the push/pull axis is broken because the drop is motile at zero force, this time due to treadmilling self-advection, $\mathcal{W}$. The RP(L) region becomes smaller with self-advection, and is pushed to the top-left (push-left, squeeze), while the LP(R) region becomes larger with with self-advection and invades the left half of the squeeze region. The RP(R) phase emerges as a thin strip between RP(L) and LP(R). DB for the polymerising drop occurs in the bottom-right (push-right, stretch) rather than in the bottom-left (push-left, stretch) as is the case for the contractile/extensile drop. The stretch region of the phase plane is populated by TM(R) and TL(R), as well as DB. The region containing TM(R) and TL(R) shares a phase boundary with both RP(R) and LP(R) determined by the second moment of $h$ as a measure of the spread of the drop, and the distribution of substrate traction. The phase boundary is indicated in the schematic Fig. \ref{fig:schematic} (c), as determined from the data in Figs. \ref{fig:profilesA} (b) and (d). Strikingly, the transition between RP(R)/LP(R) and DH(R)/PC(R) is sharp suggesting a first order transition. The behaviour of the second moment and the drop velocity across the phase boundary is shown in Fig. \ref{fig:profilesA} (e). There is a jump in both quantities going from RP(R)/LP(R) to TM(R)/TL(R). The change in drop shape induced by moving from LP(R) to TL(R) is shown \ref{fig:profilesA} (f). The drop shape changes continuously within the TM(R)/TL(R) region but changes sharply at the phase boundary. In contrast, the transition between TM(R) and TL(R) is smooth.

\begin{figure}[t!]
	\centering
	\includegraphics[width=0.5\textwidth]{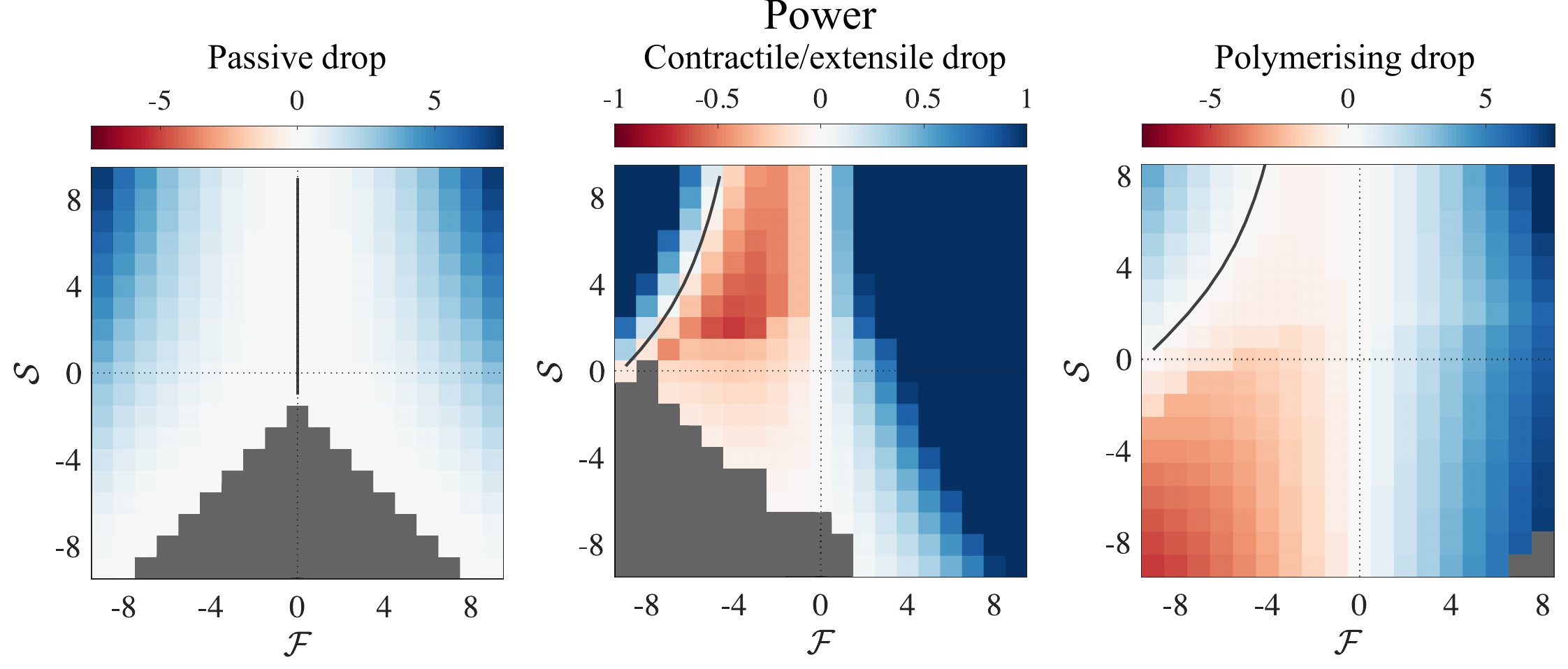}
	\caption{Left: power (net force $\times$ velocity) for a passive drop. Middle: power for an active contractile/extensile drop. Right: power for an active polymerising drop.}
   \label{fig:Power}
\end{figure}

\section{Discussion and parameter estimates}

The drastic changes in drop shape that we have observed are not associated with changes in direction of motion. Both the contractile/extensile drop and the polymerising drop remain as right-parabolas when their velocity is reversed at the RP(L)/RP(R) boundary. Stationary drop profiles are shown in Fig. \ref{fig:profiles} (h). All drops we have considered here are fastest for strong push and strong squeeze, towards the top corners of the phase plane. The contractile/extensile drop slows down dramatically on entering the DH(R) region and continues to slow down as it enters the PC(R) region and approaches DB, however it does not stop before reaching DB. In contrast, the polymerising drop locally speeds up as it approaches the TM(R) region and continues to speed up as it enters the TL(R) region and approaches DB. We can also compute the active power generated by moving the drop against applied forces, showing that when the drop moves opposite the applied force, it acts as a motile engine (see Fig. \ref{fig:Power}).

We now estimate dimensionful parameter values for our model to provide biological context and make predictions.
In our active simulations we looked at $\mathcal{A}$ between $0.3$ and $1$. Assuming a surface tension of $0.01\text{mN/m}$ \cite{gauthier2012mechanical}, a length scale of $25\mu\text{m}$, and a drop inverse aspect ratio $\epsilon = 0.1$, we can use \eqref{eq:NonDimParameters} to calculate $\tilde{\alpha}$ to be between $0.02\text{pN}/(\mu\text{m})^2$ and $0.05\text{pN}/(\mu\text{m})^2$. This is comparable to a bottom-up estimate of activity parameter $\tilde{\alpha}$ from the force produced by myosin motors cross-linking and walking across actin filaments, which we treated as rods in water. We estimated the force produced by each rod to be the Stokes drag on a rod, $F_{rod} \approx 2\pi L \eta v$, where $L$ is the length of an actin filament, which we take to be $1\mu\text{m}$, $\eta$ is the viscosity of water, which is $8.9\times10^{-4}\text{Pa s}$, and $v$ is the myosin walking speed, which we take to be $200\text{nm/s}$ \cite{howard2002mechanics}. Using these numbers we estimate the force produced by a single actomyosin bundle to be $5\text{nN}$. We take the density of actin monomers in a cell to be $100\text{mg/ml}$ \cite{janmey1991mechanical}, the mass of an actin monomer to be $42\text{kDa}$ \cite{mair1994clinical}, the volume of a cell to be $10^{-15}\text{m}^3$ \cite{hannezo2014theory}, and the number of actin monomers to per filament to be 370 \cite{lennarz2013encyclopedia}, which gives $4\times10^5$ actin filaments per cell. Multiplying this number by the Stokes drag gives a total force of $F_{cell} = 5\times10^{-10}\text{N}$. Since disordered actomyosin networks do not convert microscopic forces efficiently, we  estimate the active force scale to be $0.01-0.1 \:F_{cell}/(L_{cell}^2)$, where we have used $L_{cell} \approx 25\mu{m}$ \cite{hannezo2014theory}, so $\tilde{\alpha} \approx 0.08-0.8\text{N}/{m}^2 = 0.08-0.8\text{pN}/(\mu\text{m})^2$.

Similarly, we can calculate $\tilde{w}_0$ and $\tilde{V}$ via equation \eqref{eq:NonDimParameters} using our assumed value of surface tension, and using the viscosity of a semi-flexible polymer gel, which we take to be $3\times10^{-1}\text{Pa s}$ \cite{valberg1987magnetic}. We find $\tilde{w}_0$ between $10\text{nm/s}$ and $30\text{nm/s}$, comparable to the range of polymerisation speeds, $7\text{nm/s}$ to $170\text{nm/s}$, for motile cells \cite{carlsson2010actin}. Our model predicts cell speeds around $10\text{nm/s}$. We use values of push and squeeze ranging from $0.04\text{pN}/(\mu\text{m})^2$ to $0.4\text{pN}/(\mu\text{m})^2$, and our model predicts a stall force of around $60\text{pN}$ (assuming a contact area of $2\times 25(\mu\text{m})^2$).

\section{Conclusion}

In summary, we have studied the dynamics of a model cell (an active LC drop) on a flat surface under external forces applied at its two ends. Our phase diagrams in terms of the sum and differences of these forces and show how cell shape and motile behaviour is nonlinearly modulated by external forces. Our analysis has focused on two mechanisms of motility driven by contractile/extensile active stresses and actin polymerisation. Both mechanisms give rise to the same parabolic phases (RP(L), LP(R), RP(R)) under squeeze, and different phases under stretch: DH(R) and PC(R) for the contractile/extensile drop, and TM(R) and TL(R) for the polymerising drop. 

These unexpectedly strong shape and motility changes under applied forces should help us understand similar changes in experiment. Epithelia change from columnar (tall) to squamous (flat) shapes under tension, and there is a tradeoff between active traction with the substrate and at the apical (top) surface. Similarly, in the mesenchymal-epithelial transition from flat or humped, strongly motile mesenchymal cells slow down and become taller when squeezed in by other cells \cite{lecuit2007cell,Mol_bio_of_the_cell}. The changes in shape that we have observed here can be exploited to control behaviour of different cell phenotypes and tissue remodelling. In particular, the relation that we have derived between cell-substrate traction and forces, and between cell velocity and traction could be investigated by measuring the cell-substrate force using e.g. traction force microscopy and pushing/pulling the cell with a micropipette.

\begin{acknowledgments}
Part of this work was funded by a Leverhulme Trust Research Project Grant RPG-2016-147.
TBL acknowledges support of BrisSynBio, a BBSRC/EPSRC Advanced Synthetic Biology Research Centre (grant number BB/L01386X/1). SH acknowledges support of BBSRC grant BB/N009150/2.
\end{acknowledgments}

\bibliography{forced_drop}

\clearpage
\newpage

\begin{widetext}
\begin{center}
\textbf{\large The nonlinear motion of cells subject to external forces: Supplemental Material}
\end{center}
\end{widetext}

\setcounter{equation}{0}
\setcounter{figure}{0}
\setcounter{table}{0}
\setcounter{section}{0}
\setcounter{page}{1}
\makeatletter
\renewcommand{\theequation}{S\arabic{equation}}
\renewcommand{\thefigure}{S\arabic{figure}}
\renewcommand{\bibnumfmt}[1]{[S#1]}
\renewcommand{\citenumfont}[1]{S#1}

\onecolumngrid

\section{Deriving the height equation}

In this section we will outline the derivation of $h^{\prime\prime\prime}(\tfrac{1}{3}h^3 + l_{u}h^2) + g(h) = \mathcal{V}h$ for the active contractile drop. The derivation for the active polymerising drop is almost identical except that there is no imposed winding and the the director angle therfore scales differently, activity $\alpha$ scales differently, and there is a non-zero advection velocity in the statement of mass conservation. We begin with the force balance equation in the lubrication approximation
\begin{subequations} \label{thin_film_equations}
\begin{align}
    \partial_{\tilde{x}}\tilde{p} - \tilde{f}_{\tilde{x}} & = \eta\partial^2_{\tilde{z}} \tilde{u}_{\tilde{x}} - \tilde{\alpha}\partial_{\tilde{z}}(n_xn_z) = \partial_{\tilde{z}}\tilde{\sigma}_{\tilde{x}\tilde{z}}, \label{thin_film_equations1} \\
    \partial_{\tilde{z}}\tilde{p} & = 0, \label{thin_film_equations2}
\end{align}
\end{subequations}
where we have chosen
\begin{align}
\tilde{f}_{\tilde{x}} = \tilde{f}_{+}\delta(\tilde{x}-\tfrac{\tilde{L}}{2}) + \tilde{f}_{-}\delta(\tilde{x}+\tfrac{\tilde{L}}{2}), && \tilde{f}_{\tilde{z}} = 0.
\end{align}
For here, we will \textbf{omit the tildes}, but remember that the quantities in this section are \textbf{not} non-dimensionalised. 

\subsection{Mass conservation}

We obtain the statement of mass conservation from the kinematic boundary boundary condition $\tfrac{D}{Dt}(h(x,t)-z)$ = 0, where $\tfrac{D}{Dt} = \partial_{t} + (\mathbf{u} + w\mathbf{n})\cdot\mathbf{\nabla}$ is the material derivative. We then impose a travelling wave solution $h = h(x - Vt)$ on the kinematic condition, where $V$ is the unknown constant drop velocity, to obtain this statement of mass conservation
\begin{equation} \label{integrated_kinematic_condition}
\int_0^h (u_x + wn_x + V) \,dz = 0,
\end{equation}
where we have made the transformation $x \leftarrow x - Vt$, so that now $x$ is the centre of mass coordinate, and where where $\mathbf{u}$ is the fluid velocity inside the drop satisfying force balance and incompressibility, and $w\mathbf{n}$ describes the additional transport due to self-advection at speed $w$ of active units whose orientations are characterised by the director $\mathbf{n}$. For the active contractile drop, which is the case we outline below, $w = 0$.

The strategy from here will be to first integrate \eqref{thin_film_equations1} at $\pm L/2$ to get boundary conditions on $p(\pm L/2)$ imposed by the Dirac deltas. We will then solve \eqref{thin_film_equations1} away from $x = \pm L/2$, and apply the derived boundary conditions at $\pm L/2$. 

\subsection{External forces impose boundary conditions on pressure}

At $x=L/2$, we have
\begin{equation*}
    \int_{\tfrac{L}{2}-\Delta}^{\tfrac{L}{2}+\Delta}\partial_{x}p dx - f_{+}\int_{\tfrac{L}{2}-\Delta}^{\tfrac{L}{2}+\Delta}\delta(x-\tfrac{L}{2})dx - f_{-}\int_{\tfrac{L}{2}-\Delta}^{\tfrac{L}{2}+\Delta}\delta(x+\tfrac{L}{2})dx =  \int_{\tfrac{L}{2}-\Delta}^{\tfrac{L}{2}+\Delta}\partial_{z}\sigma_{xz}dx.
\end{equation*}
The last term on the LHS vanishes because of the definition of the delta function and the term on the RHS vanishes as we shrink the integration region ($\Delta \rightarrow 0^{+}$) because the integrand is continuous. The procedure is identical at $x=-L/2$. Thus we have
\begin{align*}
    p(\tfrac{L}{2}+\Delta) - p(\tfrac{L}{2}-\Delta) - f_{+} = 0, &&
    p(-\tfrac{L}{2}+\Delta) - p(-\tfrac{L}{2}-\Delta) - f_{-} = 0.
\end{align*}
We model the system to have a uniform pressure $\pi_{0} + \pi_{ref}$, where $\pi_{ref}$ is an arbitrary reference pressure while $\pi_{0}$ is the Laplace pressure generated by the surface tension of a passive drop. Because the external force is localised to $x=\pm L/2$ and we take the pressure to be uniform everywhere else outside the drop, it must be the case that $p(\tfrac{L}{2}+\Delta) = p(-\tfrac{L}{2}-\Delta) = \pi_{0} + \pi_{ref}$. We also define
\begin{align*}
   p(L/2) = \lim_{\Delta \to 0^{+}} p(L/2-\Delta), &&
    p(-L/2) = \lim_{\Delta \to 0^{+}} p(-L/2+\Delta),
\end{align*}
resulting in the boundary conditions
\begin{subequations} \label{pressure_BCs}
\begin{align} 
    p(\tfrac{L}{2}) & = \pi_{0} + \pi_{ref} - f_{+}, \label{pressureBCplus} \\
    p(-\tfrac{L}{2}) & = \pi_{0} + \pi_{ref} + f_{-}. \label{pressureBCminus}
\end{align}
\end{subequations}
Solving equation \eqref{thin_film_equations1} is now equivalent to solving 
\begin{equation} \label{pressure_gradient_no_force}
\partial_{x}p = \partial_{z}\sigma_{xz}
\end{equation}
with boundary conditions \eqref{pressure_BCs}.

\subsection{Pressure as a functional of drop height}
The pressure $p(x)$ inside the drop can be related to the drop height $h(x)$ using the normal component of the free surface boundary condition 
\begin{equation*}
    \bm{\sigma}\cdot\bm{m} = (\gamma\kappa - \pi_{ref})\bm{m},
\end{equation*}
\begin{equation*}
    \bm{m}\cdot\bm{\sigma}\cdot\bm{m} = \gamma\kappa - \pi_{ref},
\end{equation*}
where $\bm{m} = (1/\sqrt{1 + (h^{\prime})^2})(-h^{\prime},1)$. Using the scalings from the lubrication approximation, $\alpha \sim \epsilon^{-1}$, $p \sim \epsilon^{-2}$, $u \sim 1$, and $w \sim \epsilon$ (from incompressibility), $\sigma_{xx} = \sigma_{zz} \approx -p \sim \epsilon^{-2}$, $\sigma_{xz} \sim \epsilon^{-1}$, we can write the LHS of the above equation as
\begin{align*}
    \begin{split}
        \bm{m}\cdot\bm{\sigma}\cdot\bm{m} & = m_{x}\sigma_{xz}m_{z} + m_{z}\sigma_{xz}m_{x} + m_{z}\sigma_{zz}m_{z} + m_{x}\sigma_{xx}m_{x} \\ & \approx \frac{1}{1+(h^{\prime})^2}(-p(h^{\prime})^2 - p - 2h^{\prime}\sigma_{xz}) \\ & = -p -2\frac{h^{\prime}\sigma_{xz}}{1+(h^{\prime})^2} \\ & \approx -p.
    \end{split}
\end{align*}
Approximating $\kappa$ to leading order, we have
\begin{align*}
\begin{split}
    \kappa & = \frac{h^{\prime\prime}}{(1+(h^{\prime})^2)^{\frac{3}{2}}} \\ & \approx h^{\prime\prime}. 
\end{split}
\end{align*}
Thus
\begin{equation} \label{pressure_expression}
    p(x) = -\gamma h^{\prime\prime} + \pi_{ref}.
\end{equation}

\AI{We equate equations \eqref{pressure_expression} and \eqref{pressure_BCs} to get the boundary conditions $h^{\prime\prime}$ used to solve the height equations. This boundary condition does not depend on the reference pressure because both equations have a $\pi_{ref}$ with the same sign. The height equation also does not depend on the reference pressure because it depends on $p^{\prime}(x)$ and not $p(x)$. Therefore the problem does not depend on the reference pressure, and we formulate the problem in the main text without including it.}

\subsection{Height equation}

To get and ODE for $h(x)$ we write $u_x$ in terms of $h$ and $\partial_x p$, substitute this into mass conservation \eqref{integrated_kinematic_condition} and rearrange to get $\partial_x p$ in terms of $h$, i.e. $\partial_x p = f(h)$. We then integrate $\partial_x p = f(h)$ and use the boundary conditions \eqref{pressure_BCs} to calculate the drop velocity $V$, this turns out to be a functional of $h$, i.e $V \sim \int_{-L/2}^{L/2} f(h) \,dx$. After this, the ODE for $h$ is given by substituting \eqref{pressure_expression} into  $\partial_x p = f(h)$.

To get $u_x$ in terms of $h$ and $\partial_x p$, we integrate \eqref{pressure_gradient_no_force}, where $\sigma_{xz} = \eta\partial_z u_x - \alpha n_xn_z$ at leading order, twice with respect to $z$, using the partial slip boundary condition at the substrate and the tangential component of the free surface boundary condition (see main text) (which can be re-written as a condition on $\partial_{z}u_x(z=h)$). We then have
\begin{equation} \label{integrated_fluid_velocity}
    u_x = \frac{\alpha h}{4\pi\omega\eta}\bigg(1 - \cos\bigg({\frac{2\omega\pi z}{h}}\bigg)\bigg) + \frac{\partial_{x}p}{\eta}\bigg(\frac{z^2}{2} - h(z+l_{u})\bigg),
\end{equation}
where we have used $\theta = \omega\pi z/h$.

Substituting \eqref{integrated_fluid_velocity} into \eqref{integrated_kinematic_condition} and isolating the pressure gradient yields
\begin{equation} \label{explicit_pressure_gradient}
    \partial_{x}p = \frac{\eta(\alpha h/(4\pi\omega\eta) - V)}{\frac{1}{3}h^2 + l_{u}h},
\end{equation}
From here the strategy will be to integrate \eqref{explicit_pressure_gradient} and apply the boundary conditions \eqref{pressure_BCs} in order to determine the integration constant and the unknown drop velocity $V$. We find that the drop velocity is given by
\begin{equation} \label{drop_velocity}
    \eta V = \frac{\int_{-L/2}^{L/2} \frac{\alpha/(4\pi\omega)}{\frac{1}{3}h + l_{u}} \,dx + (f_{+} + f_{-})}{\int_{-L/2}^{L/2} \frac{1}{\frac{1}{3}h^2 + l_{u}h} \,dx}.
\end{equation}
As a sanity check, we see that the second term in the numerator vanishes when the forces are equal and opposite, which means that the drop velocity in unchanged by the forces when they cancel each other out. This is good. The drop velocity also vanishes when the activity $\alpha$ and both forces vanish, which is good. From here we can substitute \eqref{pressure_expression} into \eqref{explicit_pressure_gradient} to obtain
\begin{equation} \label{final_ODE}
    \frac{\gamma h^{\prime\prime\prime}}{\eta}\bigg(\frac{1}{3}h^3 + l_{u}h^2\bigg) + \frac{\alpha h^2}{4\pi\omega\eta} = Vh,
\end{equation}
where the drop velocity $V$ is given by \eqref{drop_velocity}. For the active polymerising drop, the same procedure applies, except we have $\omega = 0$, $\theta \sim \epsilon$, $\alpha \sim \epsilon^{-2}$, and the advection velocity $w > 0$. We set $\alpha = 0$ in the final result to get the equation in the main text.

\section{Additional phase diagrams}
\begin{figure*}[h!]
	\centering
	\includegraphics[width=1.0\textwidth]{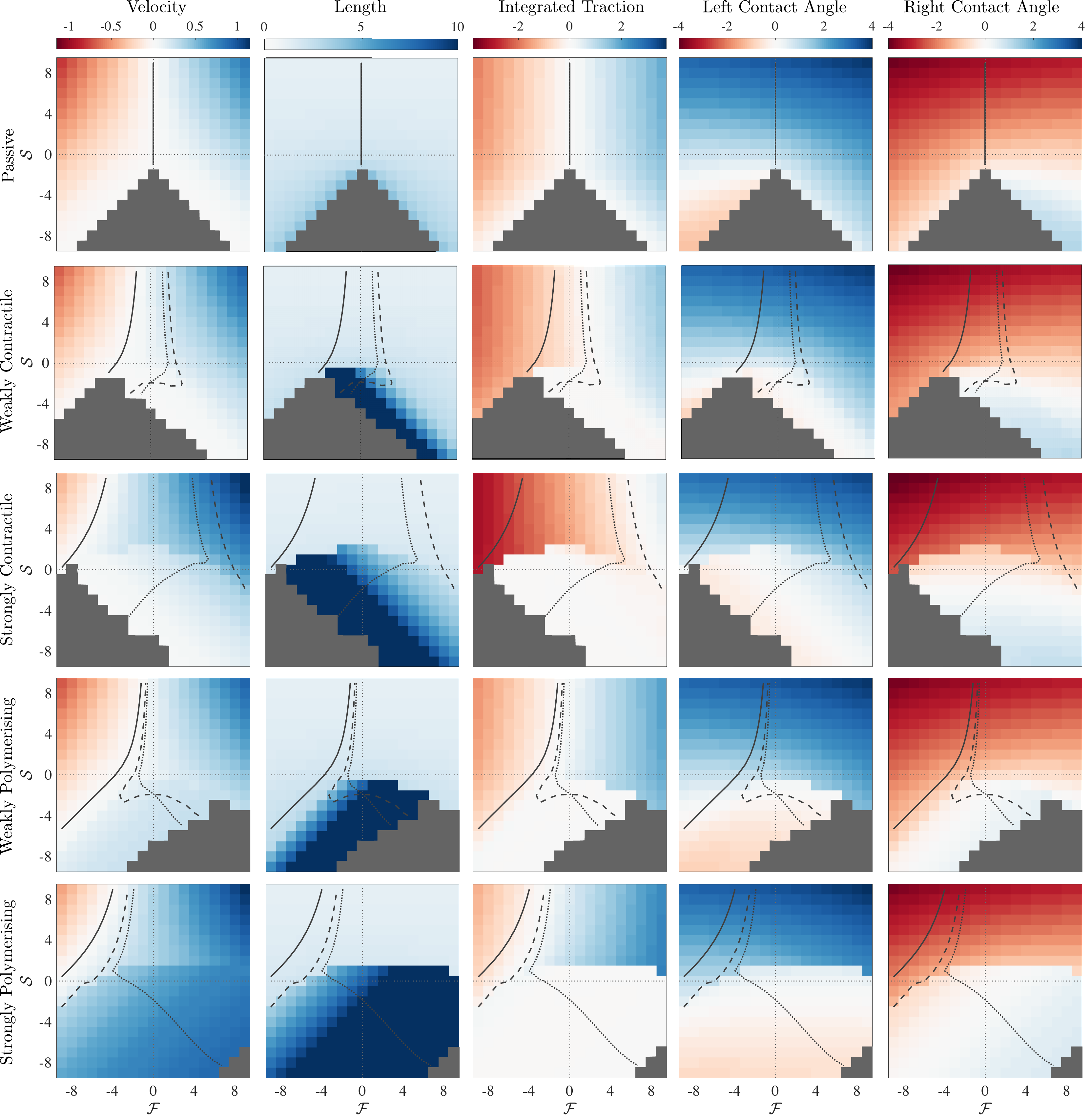}
	\caption{First row: passive drop ($\mathcal{A} = 0$), second row: weakly contractile ($\mathcal{A} = 0.3$), third row: strongly contractile ($\mathcal{A} = 1$), fourth row: weakly polymerising ($\mathcal{W} = 0.3$), fifth row: strongly polymerising ($\mathcal{W} = 1$). First column: drop velocity. Second column: drop length. Third column: raction between the drop and the substrate integrated over $70\%$ of the drop length. Fourth column: left contact angle $\phi_{-} = h^\prime(-L/2)$, fifth column: right contact angle $\phi_{+} = h^\prime(L/2)$. The solid line in each plot is an isoline corresponding to zero velocity, the dashed line is an isoline corresponding to the the 1\textsuperscript{st} moment $\mu_{1} = 0$, and the dotted line is an isoline corresponding to the drop having equal contact angles.}
   \label{fig:PhaseDiagrams}
\end{figure*}

\begin{figure*}[h!]
	\centering
	\includegraphics[width=0.750\textwidth]{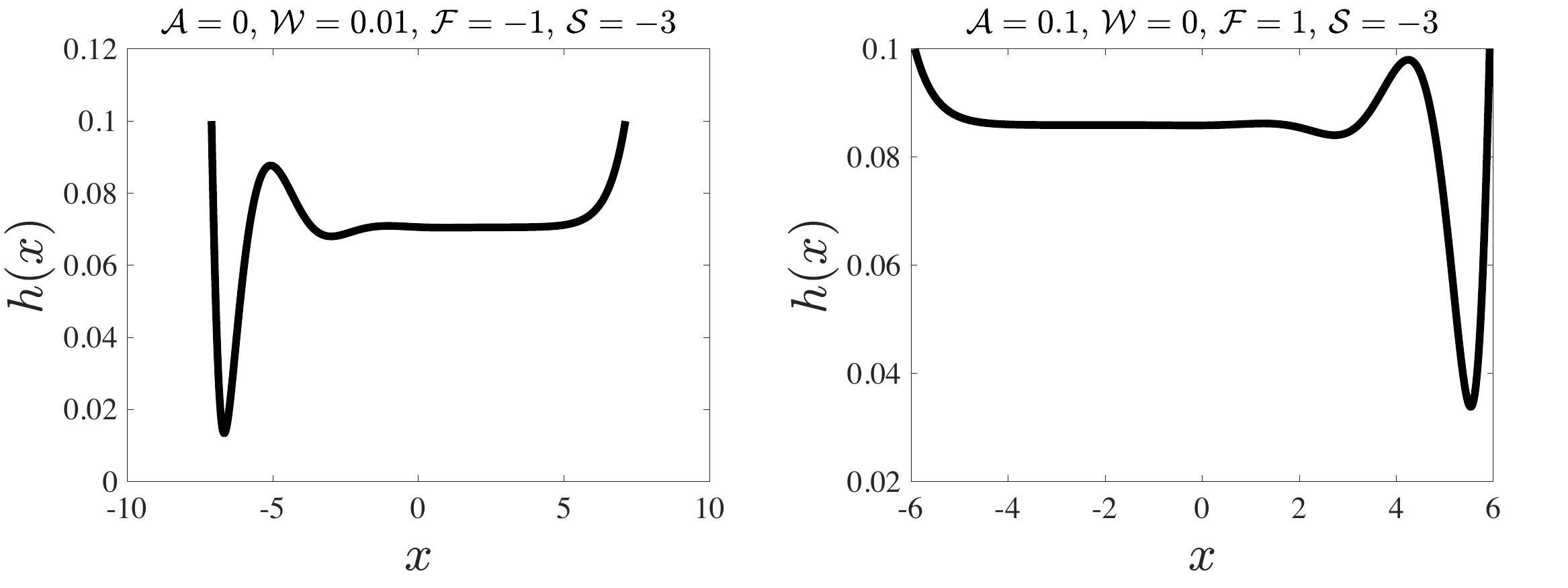}
	\caption{Left: profile for a drop with small polymerisation speed under stretch. Right: profile for a drop with small contractile stress under stretch}
   \label{fig:WeakActivityWithForces}
\end{figure*}

\section{Solving the height equation with the Crank-Nicholson method}

The equation that is fed into the Crank-Nicholson algorithm is
\begin{equation} \label{time_evolution_recast}
    \partial_{t}h + \partial_{x}(-\mathcal{V}h + h^{\prime\prime\prime}(\tfrac{1}{3}h^3 + l_{u}h^2) + g(h)) = 0,
\end{equation}
where $g(h) = \mathcal{A}h^2$ for the active contractile drop, $g(h) = \mathcal{W}l_w(1 - \exp(-h/l_w))$ for the active polymerising drop, and the drop velocity $\mathcal{V}$ is defined in the main text. Equation \eqref{time_evolution_recast} is subject to the constraints
\begin{align}
    \int_{-L/2}^{L/2}h(x) \,dx = \Omega, && h(\pm L/2) = h_{0},
\end{align}
where $\Omega$ is the dimensionless drop volume, and
\begin{align} \label{condition_on_second_derivative_nondim}
    h^{\prime\prime}(-L/2) + \pi_{0} + f_{-} = 0, && h^{\prime\prime}(L/2) + \pi_{0} - f_{+} = 0,
\end{align}
with
\begin{equation*} \label{pi_0_nondim}
    \pi_{0} = \frac{2\phi}{-\frac{3h_0}{\phi} + \sqrt{\frac{9h_{0}^2}{\phi^2} + \frac{6\Omega}{\phi}}},
\end{equation*}
where $\phi$ is the re-scaled contact angle. 

Equation \eqref{time_evolution_recast} is of the form $\partial_{t}h = I$, where $I$ represents the second term on the LHS of \eqref{time_evolution_recast}. The Crank-Nicholson scheme advances in time according to
\begin{equation} \label{CN_time_evolution}
    \frac{h_{i}^{n+1}}{\Delta t} - \frac{1}{2}I_{i}(\mathbf{X}^{n+1}) = \frac{h_{i}^{n}}{\Delta t} + \frac{1}{2}I_{i}(\mathbf{X}^{n}),
\end{equation}
where the subscripts refer to the spatial discretisation and the superscripts refer to the time discretisation. Note that in the second term on the LHS, $I_{i}(\mathbf{X}^{n+1})$, means $I$ evaluated at spatial point $i$ at time-step $n+1$. 

\subsection{Spatial discretisation}

For the discretisation we use the substitution $x=Ly$, with $L$ being the drop length, and discretise the domain $y \epsilon [-0.5,0.5]$ with uniform grid spacing $\Delta y$. The stiff term is discretised as follows
\begin{multline*}
    \partial_{x}\bigg(h^{\prime\prime}\bigg(\frac{1}{3}h^3 + l_{u}h^2\bigg)\bigg) \rightarrow  \frac{\bigg[(\frac{1}{3}h_{i}^3 + l_{u}h_{i}^2) + (\frac{1}{3}h_{i+1}^3+l_{u}h_{i+1}^2)\bigg](h_{i+2}-3h_{i+1}+3h_{i}-h_{i-1})}{2L^4\Delta y ^4} \\ -  \frac{\bigg[(\frac{1}{3}h_{i-1}^3 + l_{u}h_{i-1}^2) + (\frac{1}{3}h_{i}^3+l_{u}h_{i}^2)\bigg](h_{i+1}-3h_{i}+3h_{i-1}-h_{i-2})}{2L^4\Delta y ^4},
\end{multline*}
and all other terms are discretised as
\begin{equation*}
    \partial_{x}G(h) \rightarrow \frac{G(h_{i+1}) - G(h_{i-1})}{2L\Delta y}.
\end{equation*}

\subsection{Algorithm and boundary conditions}

Equation \eqref{CN_time_evolution} leads to a set of nonlinear algebraic equations for $\{h_{i}^{n+1}\}$, where $\{h_{i}^{n}\}$ are known, which is solved using the Matlab fsolve algorithm. For $N$ spatial grid points, define $\mathbf{X}^{n+1} = (h_{1}, ..., h_{N})^{n+1}$. The algorithm solves $\mathbf{F}(\mathbf{X}^{n+1})=\mathbf{0}$ with 
\begin{align} \label{interior_values}
    \mathbf{F}(\mathbf{X}^{n+1}) = \frac{\mathbf{X}^{n+1}}{\Delta t} - \frac{1}{2}\mathbf{I}(\mathbf{X}^{n+1}) - \frac{\mathbf{X}^n}{\Delta t} - \frac{1}{2}\mathbf{I}(\mathbf{X}^{n}) && i = 3, ..., N-2,
\end{align}
where $\mathbf{I} = (I_{1}, ..., I_{N})$, with $I_{i}$ being the spatial differential operator evaluated at spatial point $i$. The boundary conditions are implemented using grid points $1,2,N-1,N$.  The condition on the drop height at $\pm L/2$ is implemented as
\begin{subequations} \label{height_constraint}
\begin{align} 
    F_{1} = h_{1}^{n+1} - h_{0} \\ 
    F_{N} = h_{N}^{n+1} - h_{0}. 
\end{align}
\end{subequations}
The boundary conditions on the second derivative \eqref{condition_on_second_derivative_nondim} are implemented using finite difference coefficients to approximate the second derivative:
\begin{subequations} \label{second_derivative_BC2}
\begin{align}
    F_{2} = \frac{2h_{1}^{n+1}-5h_{2}^{n+1}+4h_{3}^{n+1}-h_{4}^{n+1}}{\Delta y^2} + L^2\bigg(\frac{\pi_0 + f_{-}}{\tilde{C}}\bigg), \\ 
    F_{N-1} = \frac{2h_{N}^{n+1}-5h_{N-1}^{n+1}+4h_{N-2}^{n+1}-h_{N-3}^{n+1}}{\Delta y^2} + L^2\bigg(\frac{\pi_0 - f_{+}}{\tilde{C}}\bigg).
\end{align}
\end{subequations}

Once equation \eqref{time_evolution_recast} is discretised, we use a nonlinear solver (MATLAB's ``fsolve") for the simultaneous equations $\mathbf{F}(\mathbf{X}^{n+1})=\mathbf{0}$, with the components of $\mathbf{F}$ given by \eqref{interior_values}, \eqref{height_constraint}, and \eqref{second_derivative_BC2}. We also calculate the Jacobian $\partial F_{i}/\partial_{h_{j}}$ explicitly and supply it to the nonlinear solver. The algorithm begins with a user set initial condition $\mathbf{X}^{1}$ which is chosen to satisfy the boundary conditions on drop height but not necessarily the boundary conditions on the second derivative. The drop velocity $\mathcal{V}$ as well as the drop length $L$ are calculated iteratively starting from the initial condition (the drop length is derived from the constraint that the drop has a constant volume) and plugged into the Crank-Nicholson evolution equation which is solved for $\mathbf{X}^{2}$. This process is repeated until steady state is reached i.e. the difference between $\mathbf{X}^{n+1}$ and $\mathbf{X}^{n}$ is less than some tolerance.

\section{Asymptoics for the active contractile drop} \label{asymptotics}

We expand the equation
\begin{equation} \label{non_dim_ODE}
   h^{\prime\prime\prime}(\tfrac{1}{3}h^3 + l_{u}h^2) + \mathcal{A}h^2 = \bigg(\frac{\mathcal{A}I_1 + (f_{+} + f_{-})}{I_2}\bigg)h,
\end{equation}
where $I_{1} = \int_{-L/2}^{L/2}(\tfrac{1}{3}h + l_u)^{-1}\,dx$ and $I_{2} = \int_{-L/2}^{L/2}(\tfrac{1}{3}h^2 + l_u h)^{-1}\,dx$, for small activity and small forces.

\subsection{Small forces and small activity}

We consider small perturbations to a symmetric passive drop by expanding $h(x)$ to linear order in $f_{\pm}$ and $\mathcal{A}$ around the passive solution, obtained by setting $f_{+} = f_{-} = 0$ and $\mathcal{A} = 0$ in \eqref{non_dim_ODE} and either imposing $h^{\prime\prime}(x) = \pi_{0}$ or $h^{\prime}(\tfrac{\pm L}{2}) = \mp\phi$ along with $h(\tfrac{\pm L}{2}) = h_{0}$:
\begin{equation}
   h_{p} = -\frac{\phi L}{4}\bigg(\frac{4x^2}{L^2} - 1\bigg) + h_{0}.
\end{equation}
We also expand $L$ to linear order in $f_{\pm}$ and $\mathcal{A}$:
\begin{equation}\label{expand_L}
L = L_p - f_{+}L_{+} - f_{-}L_{-} + \mathcal{A}L_{\alpha} + \cdots
\end{equation}
which leads to
\begin{equation} \label{expand_h}
\begin{split}
h(y) = & H_{0} + f_{+}\bigg(\frac{\phi L_{+}}{4}(y^2 - 1) - h_{+}\bigg) + f_{-}\bigg(\frac{\phi L_{-}}{4}(y^2 - 1) - h_{-}\bigg) \\ & + \mathcal{A}\bigg(h_{\alpha} - \frac{\phi L_{\alpha}}{4}(y^2 - 1)\bigg) + \cdots
\end{split}
\end{equation}
where $y = 2x/L$, $H_{0} = -\tfrac{\phi L_{p}}{4}(y^2 - 1) + h_{0}$, and $L_{p}$ is the length of the passive drop given by $L_{p} = \sqrt{\tfrac{6\Omega}{\phi} + \tfrac{9h_{0}^{2}}{\phi^2}} - \tfrac{3h_0}{\phi}$ for a drop of volume $\Omega$. Substituting \eqref{expand_h} into \eqref{non_dim_ODE} and keeping terms only to linear order yields three differential equations at $\mathcal{O}(f_{+})$, $\mathcal{O}(f_{-})$, and $\mathcal{O}(\mathcal{A})$ that can be integrated for $h_{\pm}$ and $h_{\alpha}$. At $\mathcal{O}(f_{+})$ we have
\begin{align} \label{linear_order_fplus}
     h_{+}^{\prime\prime\prime}(x)(\tfrac{1}{3}H_{0}^2 + l_{u}H_{0}) = -\frac{1}{I_{2}}, && h_{+}(\pm 1) = 0, && h^{\prime\prime}(1) = -\frac{L_{p}^2}{4} - \frac{\phi L_{+}}{2}.
\end{align}
For consistency, we must have also $h_{+}^{\prime\prime}(-1) = -\phi L_{+}/2$ but we cannot impose this on the equation, as there are already three boundary conditions. Fortunately, it falls out automatically because the integrals $I_{1}$ and $I_{2}$ encode both boundary conditions on $h^{\prime\prime}$. At $\mathcal{O}(f_{-})$ we have
\begin{align} \label{linear_order_fminus}
     h_{-}^{\prime\prime\prime}(x)(\tfrac{1}{3}H_{0}^2 + l_{u}H_{0}) = -\frac{1}{I_{2}}, && h_{-}(\pm 1) = 0, && h_{-}^{\prime\prime}(-1) = \frac{L_{p}^2}{4} - \frac{\phi L_{-}}{2}.
\end{align}
Again, for consistency, we must have $h_{-}^{\prime\prime}(1) = -\phi L_{-}/2$, which again falls out automatically. At $\mathcal{O}(\mathcal{A})$ we have 
\begin{align} \label{linear_order_activity}
     h_{\alpha}^{\prime\prime\prime}(x)(\tfrac{1}{3}H_{0}^2 + l_{u}H_{0}) = \frac{I_{1}}{I_{2}} - H_{0}, && h_{\alpha}(\pm 1) = 0, && h_{\alpha}^{\prime\prime}(1) = -\frac{\phi L_{\alpha}}{2}.
\end{align}
For consistency we must have $h_{\alpha}^{\prime\prime}(- 1) = -\phi L_{\alpha}/2$, which falls out automatically. To perform the integration, we use the variable $y = \tfrac{2x}{L}$. It is useful to define the following function
\begin{align*}
    G_{f}(y) = & \bigg(1+\tfrac{y}{\sqrt{\Gamma}}\bigg)^2\log{\bigg(1+\tfrac{y}{\sqrt{\Gamma}}\bigg)} - \bigg(1-\tfrac{y}{\sqrt{\Gamma}}\bigg)^2\log{\bigg(1- \tfrac{y}{\sqrt{\Gamma}}\bigg)} \\ & + \frac{1}{\beta\phi}\bigg[\bigg(\beta - \tfrac{\phi y}{\sqrt{\Gamma}}\bigg)^2\log{\bigg(\beta-\tfrac{\phi y}{\sqrt{\Gamma}}\bigg)} - \bigg(\beta + \tfrac{\phi y}{\sqrt{\Gamma}}\bigg)^2\log{\bigg(\beta +\tfrac{\phi y}{\sqrt{\Gamma}}\bigg)}\bigg],
\end{align*}
where
\begin{equation*}
\Gamma = 1 + \frac{4h_{0}}{\phi L_{p}},
\end{equation*}
and
\begin{equation*}
    \beta = \sqrt{\frac{\Gamma + \tfrac{12l_{u}}{\phi L_{p}}}{\Gamma}}.
\end{equation*}
We also need 
\begin{equation*}
G_{\alpha}(y) = \frac{\phi}{4\beta}\bigg[\bigg(\frac{\beta}{\phi} + \frac{y}{\sqrt{\Gamma}}\bigg)^2\log\bigg(\frac{\beta}{\phi} + \frac{y}{\sqrt{\Gamma}}\bigg) - \bigg(\frac{\beta}{\phi} - \frac{y}{\sqrt{\Gamma}}\bigg)^2\log\bigg(\frac{\beta}{\phi} - \frac{y}{\sqrt{\Gamma}}\bigg)\bigg].
\end{equation*}
In terms of these functions, the leading order contributions to the drop height are 
\begin{subequations}
\begin{align}
& h_{+}(y) = \frac{-L_{p}^2(1 + \tfrac{4h_{0}}{\phi L_p})}{8G_{f}^{\prime\prime}(1)}\bigg[G_{f}(y) + \tfrac{1}{2}G_{f}^{\prime\prime}(1))\bigg(1 + \frac{4\phi} L_{+}{L_{p}^2}\bigg)y^2 - \sqrt{1 + \tfrac{4h_{0}}{\phi L_{p}}}G_{f}(1)y - \frac{G_{f}^{\prime\prime}(1)(1 + 4\phi L_{+}/L_{p}^2)}{2(1+4h_{0}/\phi L_{p})}\bigg], \\
& h_{-}(y) = \frac{-L_{p}^2(1 + \tfrac{4h_{0}}{\phi L_{p}})}{8G_{f}^{\prime\prime}(1)}\bigg[G_{f}(y) - \tfrac{1}{2}G_{f}^{\prime\prime}(1)\bigg(1 - \frac{4\phi L_{-}}{L_{p}^2}\bigg)y^2 - \sqrt{1 + \tfrac{4h_{0}}{\phi L_{p}}}G_{f}(1)y + \frac{G_{f}^{\prime\prime}(1)(1 - 4\phi L_{-}/L_{p}^2)}{2(1+4h_{0}/\phi L_{p})}\bigg], \\
& h_{\alpha}(y) = \frac{3L_{p}^2\sqrt{1+\tfrac{4h_{0}}{\phi L_{p}}}}{2\phi}\bigg[\frac{G_{\alpha}^{\prime\prime}(1)}{G_{f}^{\prime\prime}(1)}G_{f}(y)- G_{\alpha}(y) + \sqrt{1+\tfrac{4h_{0}}{\phi L_{p}}}\bigg(G_{\alpha}(1) -\frac{G_{\alpha}^{\prime\prime}(1)}{G_{f}^{\prime\prime}(1)}G_{f}(1)\bigg)y\bigg],
\end{align}
\end{subequations}
where
\begin{subequations}
\begin{align}
& L_{+} = \frac{-L_{p}^3}{24}\bigg(\frac{\phi L_{p}}{2}+ h_{0}\bigg)^{-1}, \\
& L_{-} = \frac{L_{p}^3}{24}\bigg(\frac{\phi L_{p}}{2} + h_{0}\bigg)^{-1}, \\
& L_{\alpha} = 0.
\end{align}
\end{subequations}

\section{Hints of bi-stability}

\begin{figure*}[h!]
	\centering
	\includegraphics[scale = 0.215]{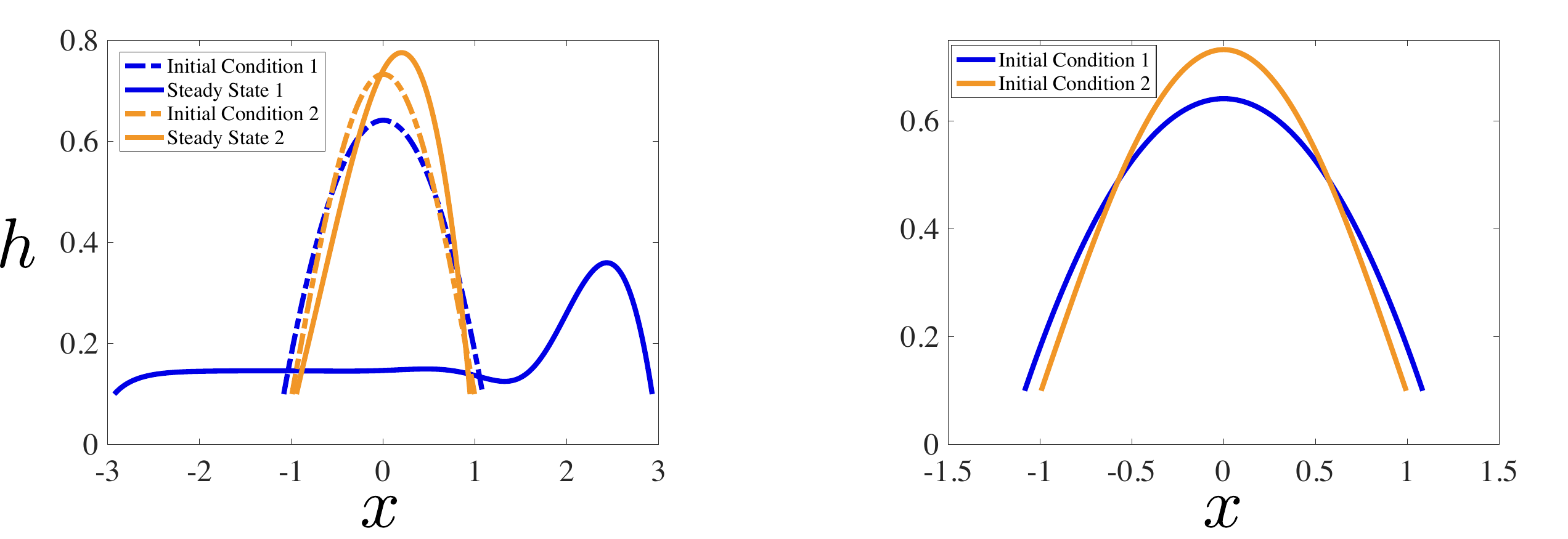}
	\caption{Left: Initial height profiles (dashed lines) and steady state profiles (solid lines) for  Right: Only the initial height profiles. Here $\mathcal{A} = 1$, $\mathcal{F} = -2$, and $\mathcal{S} = 2$.}
   \label{fig:changeIC}
\end{figure*}

Both the active contractile drop and active polymerising drop can produce different steady state drop profiles in the iterative numerical scheme, starting from different height profiles with the same activity and applied forces. This is shown in figure \ref{fig:changeIC} for the active contractile drop at the RP(R)/DH(R) boundary. For the active polymerising drop, this behaviour occurs at the LP(R)/TM(R) boundary.

\end{document}